# Accurate prediction of chemical short-range order and its effect on thermodynamic, structural, and electronic properties of disordered alloys: exemplified in Cu$_3$Au

Will Morris[a], Duane D. Johnson[a,b], and Prashant Singh[a,*]

[a] Ames National Laboratory, U.S. Department of Energy, Iowa State University, Ames, Iowa 50011, USA
[b] Department of Materials Science & Engineering, Iowa State University, Ames, IA 50011, USA

**Abstract**

Electronic-structure methods based on density-functional theory (DFT) were used to directly quantify the effect of chemical short-range order (SRO) on thermodynamic, structural, and electronic properties of archetypal face-centered-cubic (fcc) Cu$_3$Au alloy. We show that SRO can be tuned to alter bonding and lattice dynamics (i.e., phonons) and detail how these properties are changed with SRO. Thermodynamically favorable SRO improves phase stability of Cu$_3$Au from -0.0343 eV-atom$^{-1}$ to –0.0682 eV-atom$^{-1}$. We use DFT-based linear-response theory to predict SRO and its electronic origin, and accurately estimate the transition temperature, ordering instability (L1$_2$), and Warren-Cowley SRO parameters, observed in experiments. The accurate prediction of real-space SRO gives an edge over computationally and resource intensive approaches such as Monte-Carlo methods or experiments, which will enable large-scale molecular dynamic simulations by providing supercells with optimized SRO. We also analyze phonon dispersion and estimate the vibrational entropy changes in Cu$_3$Au (from 9k$_B$ at 300 K to 6k$_B$ at 100 K). We establish from SRO analysis that exclusion of chemical interactions may lead to a skewed view of true properties in chemically complex alloys. The first-principles methods described here are applicable to any arbitrary complex solid-solution alloys, including multi-principal-element alloys, so hold promise for designing technologically useful materials.

**Introduction**

Thermally induced short-range order (SRO) in a disordered phase is related to atomic-scale chemical interactions that manifests as infinitesimal chemical fluctuations [1,2]. The SRO has been widely measured for binary Fe-Cr, Ni-Cr, and Cu-Au alloys using diffuse scattering experiments [1-6], and found to have significant impact on electronic, structural, magnetic, and mechanical properties. Gerold *et al.* [7] argued that planarity of dislocation gliding was primarily triggered by SRO in face-centered-cubic (fcc) metals, while Han *et al.* [8] found the SRO in Cu-Mn alloys drive anomalous recovery of work-hardening rates by activating planar slip. Recently, both theory and experiment have showcased the explicit role of SRO in chemically complex materials, controlling electronic [9-11], mechanical [12,13], thermoelectric [14,15], and transport [16] properties. Recently, a breakthrough in direct observations of SRO in multi-component alloys [12,17,18] has confirmed prior theoretical predictions [9,19,20] and reignited much-needed discussions on this topic. Undoubtedly, the explicit demonstration of SRO and its role on alloy properties will lead to further investigation on this challenging topic especially involving underlying chemical and structural complexity, including in complex solid solution alloys, like high-entropy alloys.

Here, we choose a simple archetypal system, fcc Cu$_3$Au, a classic example in the solid-solution alloy theory, where temperature dependence of SRO has been measured [6,21-24]. As a prototype, Cu$_3$Au is often used to investigate the difference of physical properties between the ordered and disordered phases, both experimentally and theoretically [25-31]. We focused on understanding SRO and its effect on

Corresponding author: psingh84@ameslab.gov

thermodynamic, structural, electronic, and mechanical properties using first-principles density-functional theory (DFT) methods. Once SRO parameters are determined directly by linear response, the disordered and SRO in fcc Cu$_3$Au were then modeled using *Super-Cell Random Approximates* (SCRAPs) [32]. The degree of SRO lowers the mixing energy in fcc Cu$_3$Au due to increased preference of unlike atomic pairs compared to a random solid solution (zero SRO). A systematic investigation of SRO tunability was performed to understand the effect of SRO on bonding characteristics, lattice dynamical properties (i.e., phonons), and electronic density of states. We showcase the ability of two electronic-structure methods to capture chemical correlations (SRO), and reveal its effects on thermodynamics, electronic and structural properties.

**Computational Methods**

*SCRAPs*: We used first-principles DFT as implemented in *Vienna Ab-initio Simulation Package* (VASP) [33,34] for geometrical optimization (i.e., volume and atomic positions) and charge self-consistency. The generalized-gradient approximation (GGA) of Perdew, Burke and Ernzerhof (PBE) was employed [35] with a plane-wave cut-off energy of 520 eV. The choice of PBE over LDA or meta-GGA [36,37] functionals is based on the work of Söderling *et al.* [38] and Giese *et al.* [39] that establish the effectiveness of GGA functionals. A 108-atom SCRAP (composed of $3 \times 3 \times 3$ fcc unit 4-atom cubes) was optimized with energy and force convergence criteria of $10^{-6}$ eV and $10^{-6}$ eV/Å, respectively. For Brillouin-zone integrations, a Γ-centered Monkhorst-Pack *k*-mesh was used for structural-optimization ($3 \times 3 \times 3$) and charge self-consistency ($5 \times 5 \times 5$) [40].

Density-functional perturbation theory (DFPT) was used to construct the force-constant matrix needed to calculate lattice dynamics [41-44], employing finite displacements of 0.03 Å along x-, y-, and z-direction for each atom in the supercell. The phonon dispersion was calculated along high-symmetry directions in the fcc Brillouin zone [45]. The phonon calculations were done on a 108-atom SCRAPs using a 5×5×5 Γ-centered *k*-mesh and an energy-convergence criteria of $10^{-7}$ eV. The bonding analysis was performed using projected crystal-orbital Hamilton population (pCOHP), as implemented in Local Orbital Basis Suite Towards Electronic-Structure Reconstruction (LOBSTER) code [46]. The off-site densities-of-states in COHP are weighted by respective Hamilton matrix elements to reveal bonding, nonbonding, and antibonding types in the given optimized crystalline SCRAP [46-48] that mimic the disordered alloy.

*SRO linear-response theory*: SRO in disordered phase describes, similar to displacement fluctuations in phonons, the compositional fluctuations based on the DFT grand-potential and underlying electronic structure [49-53]. The DFT-based grand-potential is analytically expanded to *second-order* in site-occupation probabilities (i.e., concentrations) $c_{Cu}^I$ and $c_{Au}^j$ at lattice sites *i,j* to calculate chemical stability matrix $S_{Cu-Au}^{(2)}(\mathbf{k}; T)$ [9,48]. The $S_{Cu-Au}^{(2)}(\mathbf{k}; T)$ is pair-interchange energies for Cu-Au pairs (not a pair interaction), which was calculated for all wavevector **k** in the fcc Brillouin zone. The Warren-Cowley SRO parameters $\alpha_{Cu-Au}(\mathbf{k}; T)$ are given by an inverse relation to $S_{Cu-Au}^{(2)}(\mathbf{k}; T)$, obtained from DFT [9,10,51]. Real-space $\alpha_{\mu\nu}^{ij}$ was calculated by inverse Fourier transform over real-space shells, as done experimentally. Notably, the eigenvectors of $S_{\mu\nu}^{(2)}(\mathbf{k}; T)$ reveal the dominant ($\mathbf{k} = \mathbf{k}_o$) chemical ordering (Fourier) mode that lowers the free energy for correlated fluctuations of $c_\mu^I$ and $c_\nu^j$ between sites [9,49]. For a dominant $\mathbf{k}_o$, the SRO diverges at the spinodal temperature ($T_{sp}$) due to the absolute instability at $\mathbf{k}_o$, i.e., $\alpha_{Cu-Au}^{-1}(\mathbf{k}_o; T_{sp})=0$, providing an estimate for the order-disorder temperature [9,49], or the miscibility temperature (if $\mathbf{k_o} = 0$). Example applications to binary, ternary, and multi-principal-element alloys have yielded excellent predictions on design (e.g., [9]), including for vacancy-mediated ordering [11].

**Generalized Concentration (Fourier) Wave**: The partial long-range order reflected in the SRO is interpreted by Fourier analysis, where $\mathbf{e}_G$ (normal modes in Gibbs' space) is obtained from $S^{(2)}$ stability matrix [9]. In homogeneously substitutional solid solutions, the occupation $n(\mathbf{r}_i)$ probabilities at site $\mathbf{r}_i$ are identical to composition $c$ $for$ $all$ $i$. In an ordered phase, it depends on the type of order. For N-component alloys, where all sites are represented by the same Bravais lattice, the occupations can be expanded in a Fourier series (concentration wave) and written in terms of normal modes as [19]:

$$\begin{bmatrix} n^1(\mathbf{r}) \\ n^2(\mathbf{r}) \\ n^3(\mathbf{r}) \\ \vdots \\ n^{N-1}(\mathbf{r}) \end{bmatrix} = \begin{bmatrix} c^1 \\ c^2 \\ c^3 \\ \vdots \\ c^{N-1} \end{bmatrix} + \Sigma_{s,\sigma} \eta_\sigma^s \begin{bmatrix} e_\sigma^1(\mathbf{k}_s) \\ e_\sigma^2(\mathbf{k}_s) \\ e_\sigma^3(\mathbf{k}_s) \\ \vdots \\ e_\sigma^{N-1}(\mathbf{k}_s) \end{bmatrix} \times \Sigma_{j_s} \gamma_\sigma(\mathbf{k}_{j_s}) e^{i\mathbf{k}_{j_s}\cdot\mathbf{r}} \quad . \quad \text{Eq. 1}$$

For a given atomic position $\mathbf{r}$, $\mathbf{c}$ is the composition vector of (N–1)$^{th}$ component, relative to "host" element N. Sums run over the star *s* (inequivalent wavevectors defining the order), $\sigma$ (eigenvector branch of the free-energy quadric), and *j_s* (equivalent wavevectors in star *s*). The other quantities are: long-range order (LRO) parameter $\eta_\sigma^s$ (defined: $0 \le \eta(T) \le 1$) for the $\sigma^{th}$ branch and *s*-star; $\mathbf{e}_\sigma(\mathbf{k})$ eigenvector of the normal concentration mode for branch $\sigma$; and symmetry coefficients $\gamma_\sigma(\mathbf{k}_{j_s})$ found by normalization condition and lattice geometry. Note that the elements of vectors **n** and **c** must sum to 1 to conserve probability, i.e., $\sum_{i=1}^{N} c_i = 1$, so the N$^{th}$ element information is given by the sum of the first (N – 1).

**Results and Discussion**

At temperatures above spinodal temperature ($T_{sp}$), alloys often form solid solutions. Well above $T_{sp}$, the solid solution is approaches a homogeneously random. However, as the temperature approaches $T_{sp}$ the atoms in a solid solution exhibit correlated (not fully random), rather they show some degree of SRO from preferential chemical pair-interchange energies that lower the free energy [1,6,48]. The Warren-Cowley SRO parameters $\alpha_{\mu\nu}^{ij}$ in a disordered alloy have a finite range given analytically by [1,32]

$$-\frac{\min(c_\mu, c_\nu)^2}{c_\mu c_\nu} \le \alpha_{\mu\nu}^{ij} \le +1 \qquad \text{Eq. (2)}$$

where $(c_\mu, c_\nu)$ are site probabilities of $\mu$ ($\nu$) atom at site $i$ ($j$). In terms of SRO, the pair probabilities in Eq. (2) are $P_{\mu\nu}^{ij} = c_\mu^i c_\nu^j (1 - \alpha_{\mu\nu}^{ij})$. The SRO in real-space [$(i-j)$ denotes a neighbor distance (shell)] can have three possibilities, (i) $\alpha_{\mu\nu}^{ij} < 0$ (negative; ordering-type SRO), (ii) $\alpha_{\mu\nu}^{ij} = 0$ (fully random, zero SRO), and (iii) $\alpha_{\mu\nu}^{ij} > 0$ (positive; clustering-type SRO). The free energy of a high-temperature disordered phase (i.e., $\alpha_{\mu\nu}^{ij} = 0$) will be higher due to absence of favored chemical interaction among alloying elements. From this definition, tuning degree of SRO can change both phase stability and structural properties.

The deviation from disorder phase determines the degree of order, i.e., SRO, in chemically complex alloys. The 108-atom SCRAPs [32] was optimized with specified experimental SRO up to three shells [6,21] to mimic a disorder ($\alpha_{\mu\nu}^{ij}$=0 at 823 K) and SRO ($\alpha_{\mu\nu}^{ij} < 0$ at 723 K and 678 K) as shown in **Fig. 1.**

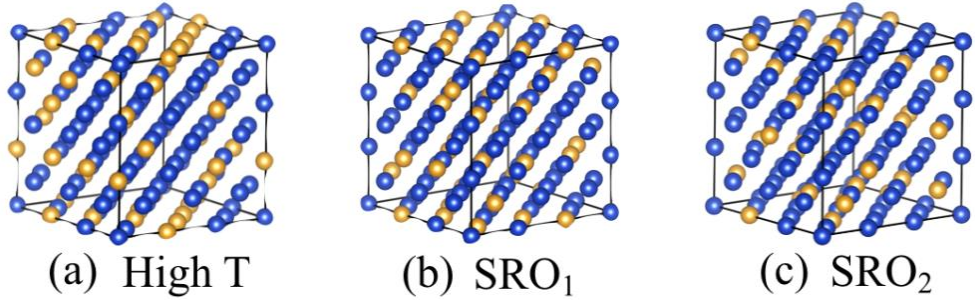

**Fig. 1**. A 108-atom SCRAPs with optimized SRO for fcc $Cu_3Au$ at (a) high-T (disorder; 823 K), $SRO_1$ (723 K), and $SRO_2$ (678 K) above order/disorder [**6**].

Next we performed inverse Fourier transform to calculate real-space Warren-Cowley SRO parameters $\alpha_{lmn}$ (*lmn* denotes real-space shells) using our DFT-SRO method. The comparison with experiments in **Table 1** shows good agreement with famous experiments by Moss et al. [**6**]. Similar to experiments, our calculations show 1$^{st}$ and 2$^{nd}$ shells as the most dominant real-space SRO modes. The ordering tendency increases with decreasing temperature. The reproducibility of SRO experiments for fcc $Cu_3Au$ establishes the ability of our method [**9**] to accurately predict real-space SROs for generating large supercells required for deformation or mechanical proeprty study within molecular dynamic framework.

**Table 1**. The DFT-SRO parameters compared with experiments [**6**]. The central site $\alpha_{000} = 1$ for DFT by construction, whereas experiment has error that provides an estimate of error in observed parameters.

| Shell (*lmn*) | Real-space SRO [$\alpha_{lmn}$] | | | |
|---|---|---|---|---|
| | DFT | | Measured [6] | |
| | 723 K | 678 K | 723 K | 678 K |
| **000** | 1.000 | 1.000 | 1.140 | 1.280 |
| **110** | -0.158 | -0.193 | -0.195 | -0.218 |
| **200** | +0.240 | +0.263 | +0.215 | +0.286 |
| **211** | +0.008 | -0.023 | +0.003 | -0.012 |

Furthermore, predicting chemical instabilities in disordered alloys, i.e., ordering modes and their origin, are of great practical and fundamental interest [**9**]. The ordering can be determined by assessing entropy (favoring disorder) and chemical interactions (favoring LRO), driven by underlying electronic effects, such as hybridization, band-filling, Kohn anomalies or Fermi-surface nesting. Thus, we analyzed SRO in $Cu_3Au$ via DFT-based linear-response SRO method [**49**]. In **Fig. 2a,b,** we plot Warren-Cowley $\alpha_{Cu-Au}(\mathbf{k}; T)$ and chemical stability $S^{(2)}_{Cu-Au}(\mathbf{k}; T)$ at T=1.15$T_{sp}$, respectively. The dominant SRO peaks for fcc $Cu_3Au$ at wavevector $\mathbf{k_o}=\mathbf{X}=(100)$ is indicative of $L1_2$-type ordering [**54-57**].

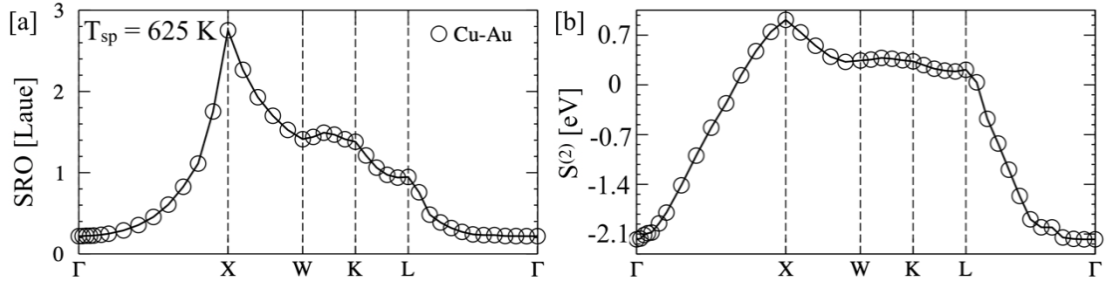

**Fig. 2.** (a) Warren-Cowley SRO parameter, and (b) chemical stability matrix for fcc $Cu_3Au$ at $T=1.15T_{sp}$ with estimated $T_{sp}$ of 625 K. The SRO peak at X-point ($k$=100) in fcc Brillouin zone is signature of $L1_2$ ordering, as observed experimentally [54-57].

The SRO is calculated from an analytic second variation (obtained exactly) of the DFT Grand Potential, here performed in "band-energy-only" approximation [9], as second variation of doubling-counting terms are small by Anderson's force theorem in the high-T solid-solution phase. When symmetry is broken below an order-disorder boundary, however, then those terms would contribute – but the linear-response is not done below spinodal temperature. Nonetheless, we can then address the partial LRO as more properly handled in the full scf-KKR-CPA code by solving analytical concentration wave equation in Eq. (1) as input to the expected partially LRO state below the spinodal temperature.

**Concentration wave analysis for binary $Cu_3Au$**: The fcc $Cu_3Au$ shows SRO peak at $\mathbf{k}_1=X$, which signify $L1_2$-type partially order. The occupation at each ordering sublattice for $Cu_3Au$ can be written as a joint Fourier wave making use of Eq. 1 that addresses $L1_2$ ordering, i.e.,

$$n(\mathbf{r}) = [n^{Au}(\mathbf{r})] = [0.25] + \eta_1(T) \cdot \mathbf{e}(\mathbf{k}_1) \cdot \gamma_1 e^{i\mathbf{k}_o \cdot \mathbf{r}} \qquad \text{Eq. (3)}$$

Using Eq. 2, the expected **$L1_2$** LRO state occupation variables are given by $\mathbf{k}_o = X = (100)$ as

$$[n^{Au}(\mathbf{r})] = [0.25] + \frac{1}{4}(-1)\,\eta_1(T)\,e^{i(100) \cdot \mathbf{r}} \qquad \text{Eq. (3a)}$$

Here $\mathbf{e}(X) \approx -1.0$ and $\gamma = \frac{1}{4}$. As site probabilities must sum to 1, $[n^{Cu}(\mathbf{r})] = 1 - [n^{Au}(\mathbf{r})]$. From this concentration wave, we find occupations versus $\eta_1(T)$ at cube corner $\mathbf{r}=(000)$ or face-center sites $\left(\frac{1}{2}\frac{1}{2}0\right)$, occupied by Au and Cu, respectively. Note that the LRO parameter $\eta_1(T)$ is a statistical variable that can be obtained by Monte Carlo simulations or diffraction experiments, however, not from SRO; nonetheless, we can ask what the maximum LRO allowable before Eq. 2a is invalid, which provides an estimate of the partial-LRO permissible.

Thus, the DFT-based linear-response SRO estimates the fcc-to-$L1_2$ transition temperature (625 K) for $Cu_3Au$ is in good agreement with experimental order-disorder temperature (663 K) [58]. Furthermore, we used direct DFT energies to estimate the order-disorder temperature ($T_c = 619$ K) using mean-field estimate $k_B T_c \approx \Delta E = [E_f(dis) - E_f(L1_2)]$ [59], where $E_f$ is the formation energies of competing fcc and fully ordered phases, i.e., $E_f(dis) = -0.0343$ $eV$-$atom^{-1}$ and $E_f(L1_2) = -0.0876$ $eV$-$atom^{-1}$. Notably, the DFT-SRO predicted $T_{sp}$ (625 K) is in very consistent with both experiment (663 K) and direct DFT (619 K), which can be improved by including the effect of phonon entropy [59].

The **k**-space representation of SRO, like phonons, shows the stability of ordering modes, which more clearly manifests correlations from underlying electronic effects in real space. Hence, we inverse-Fourier transform the Cu₃Au **k**-space pair-interchange energies to real-space using the relation:

$$S^{(2)}_{Cu-Au}(k) = S^{(2)}_{Cu-Au,0} + \sum_{i \in n_s} S^{(2)}_{Cu-Au,n_s} e^{i\mathbf{k}\cdot\mathbf{R}_i} \qquad \text{Eq (4)}$$

where, $n_s$ is the neighbor shell number and $R_i$ is the atomic coordinate in *i*-th shell. The real-space pair-interchange energies dictate the thermodynamic behavior that reveals the nature and range of interactions [**9**]. The $S^{(2)}_{n_s}$ was calculated using Eq. (4) at 675 K and 725 K [above the spinodal ($T_{sp}$=625 K)] over 15 nearest-neighbor shells ( **Fig. 3**). The pair-correlations are dominated by the first three-neighbors shells. $S^{(2)}_{Cu-Au}$ decrease with increasing shell number. This suggests that the origin of ordering is predominantly from band hybridization and filling (discussed below), with weak oscillation possibly from Fermi surface effects, as known in CuAu [**60**].

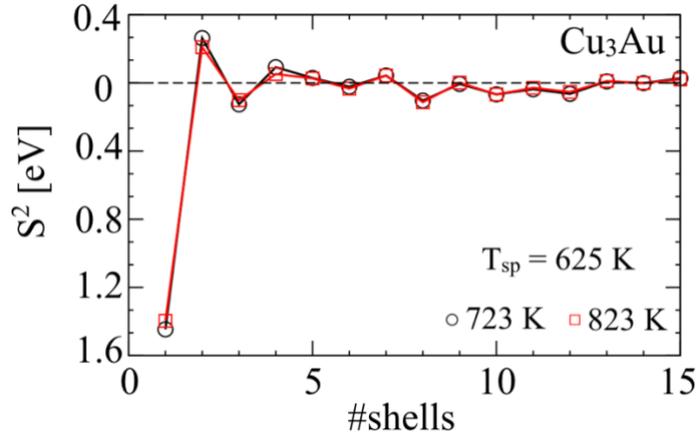

**Fig. 3.** Real-space pair-interchange energies to 15 shells for Cu₃Au, calculated at 1.15 $T_{sp}$ ($T_{sp}$ = 625 K).

We want to emphasize, here, that Greens' function based DFT KKR-CPA method directly includes competing electronic and alloying effects, i.e., band-filling, Fermi-surface nesting, van Hove singularities, hybridization, etc., which includes all relevant physics in the SRO, but which are clearly not included in model Hamiltonians. Besides this, all other approximations in SRO method arise from DFT, e.g., exchange-correlation functional. These are fully detailed, and the results are clear and compelling. Importantly, results in mean-field theory should obey to zeroth-order susceptibility sum-rule (i.e., optical theorem, or intensity sum-rule) that is included correctly in the diagonal portion to our **k**-space susceptibility [**9**], by an Onsager cavity-field correction. Notably, if a mean-field theory model Hamiltonians includes this simple correction, the phase boundary topology is completely corrected to that found by Monte Carlo [**61**]. Again, if we stay above the spinodal temperature, the linear-response SRO has no issue with phase boundary details. Besides this, as the theory is a first-principles version of Landau's theory of continuous (i.e., second order) phase transitions (relative to the highest-symmetry "ideal" disordered state) permitting solution in **k**-space, it provides the order-parameter and compares all possible wavevectors "**k**" associated with the underlying Bravais lattice. As with phonon linear response, relative to the "ideal" lattice, both short-range order (modes) is obtained, with an estimate of the long-range ordered distribution.

*Thermodynamic properties:* The lattice constants, volume, formation energy $E_f$, and DOS ($E_{Fermi}$) were calculated using DFT for three SRO cases (**Fig. 1**), and tabulated in **Table 2**, discussed later. We found that even a small change in lattice constant (0.4%) and volume (1.1%) has a strong impact on phase stability ($E_f$). For example, $E_f$ in disorder phase (–0.0343 *eV-atom$^{-1}$*; no SRO) is higher than with SRO$_1$ (-0.0637 *eV-atom$^{-1}$*) and SRO$_2$ (-0.0682 *eV-atom$^{-1}$*). The improved energy stability of solid-solution phase is attributed to increased electronic excitation arising from thermally induced SRO. The increased SRO energetically favors preferential elemental arrangement in an alloy.

**Table 2**. DFT-calculated lattice constant (Å), volume (Å$^3$-atom$^{-1}$), formation energy (E$_{form}$; eV-atom$^{-1}$), and density of states (States-eV$^{-1}$-atom$^{-1}$) for three SRO cases in **Fig. 1**.

| System [Cu$_3$Au] | $a_{lat}$ Å | V Å$^3$-atom$^{-1}$ | $E_{form}$ eV-atom$^{-1}$ | DOS $[E_F - 5\ eV]$ States-eV$^{-1}$-atom$^{-1}$ |
|---|---|---|---|---|
| High-T | 3.802 | 13.739 | -0.0343 | 1.40 |
| SRO$_1$ | 3.791 | 13.617 | -0.0637 | 1.75 |
| SRO$_2$ | 3.788 | 13.597 | -0.0682 | 1.90 |

*Structural properties:* To understand observed changes in stability of fcc Cu$_3$Au, we investigated changes to nearest-neighbor occupation and average bond-length with increasing SRO, as listed in **Table 3**. As expected, the number of Cu-Au near neighbors increased significantly (29.4%) with increase in SRO, while like-pairs decreased for Cu-Cu (11.7%) and Au-Au (100%), as required by pair probability sum rule. Moreover, a more uniform distribution of Cu-Au pairs was found with increased SRO. To further clarify, the pairs represented in **Table 3** shows the change in probability of like pairs Cu-Cu and Au-Au to occupy nearest-neighbor sites. Notably, the number of like pairs is decreased with increasing degree of SRO.

**Table 3**. The change in nearest-neighbor (NN) pair distribution and average bond-length (BL$_{avg}$) with increasing SRO in fcc Cu$_3$Au.

|  | NN pairs | | | Average bong length [Å] | | |
|---|---|---|---|---|---|---|
|  | Cu-Cu | Au-Au | Cu-Au | Cu-Cu | Au-Au | Cu-Au |
| High-T | 410 | 17 | 221 | 2.605 | 2.740 | 2.673 |
| SRO$_1$ | 375 | 1 | 272 | 2.625 | 2.748 | 2.646 |
| SRO$_2$ | 362 | 0 | 286 | 2.627 | N/A | 2.645 |
| % Change | -11.7% | -100% | 29.4% | 0.85 | N/A | -1.02 |

A random distribution (high-T) of Cu/Au in fcc Cu$_3$Au would produce a relatively large local lattice and bond distortion due to atomic-size differences and change in chemical (Cu/Au) interactions. This is also reflected through Au-Au (2.740 Å), Cu-Au (2.673 Å), and Cu-Cu (2.605 Å) bond anisotropy of the disordered phase. As follows for ordering-type SRO, the neighbor analysis indicates increased probability for Au and Cu to be a neighbor for SRO$_1$ and SRO$_2$, see **Table 3**. The uniform distribution of unlike pairs is expected to reduce both bond anisotropy and local lattice distortion compared to disordered phase. Expectedly, Cu-Au (2.625 Å) and Au-Au (2.646 Å) bond lengths become more isotropic with increasing SRO, see **Table 3** and **Fig. 4a**. To be more specific, the homoatomic pairs, i.e., Cu-Cu and Au-Au, are mostly elongated, while the unlike pairs of Cu-Au show compressed bonds. This is also expected as it is more electronically favorable for Cu to sit around Au due to increased degree of SRO (and *d*-band hybridization). A reduced average bond-length of unlike pairs agrees with reduced lattice constants ($a_{high-T} > a_{SRO1} > a_{SRO2}$) and similar for volume in **Table 2**.

The fully disordered phase has zero SRO, i.e., $\alpha_{\mu\nu}^{ij} = 0$ in Eq. (2). Then, bond anisotropy in Cu$_3$Au is possibly driven by large lattice mismatch between Cu (1.28 Å)/Au (1.44 Å), which will be the main stabilizing factor of disordered phase. In **Fig. 4a**, we plot the distribution of average bond-length for Cu [atoms 1-81] and Au [atoms 82-108] for high-T, SRO$_1$ and SRO$_2$ cases.

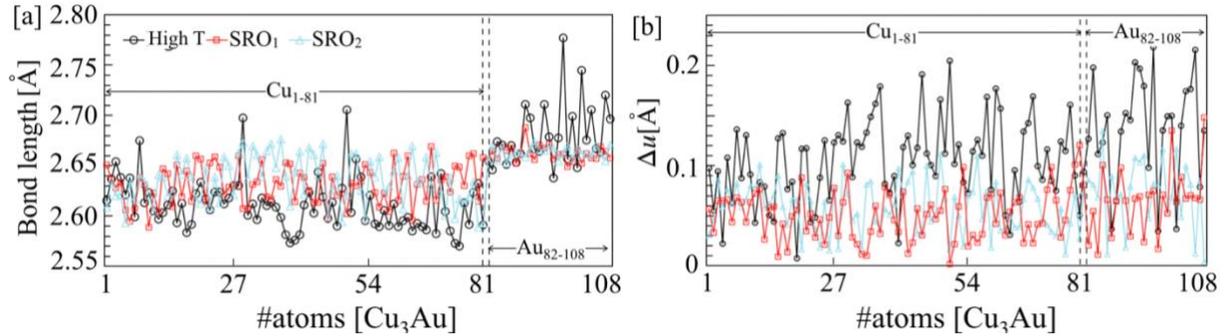

**Fig. 4**. For fcc Cu$_3$Au, (a) bond length and (b) lattice distortion ($\Delta u$) with SRO in 108-atom SCRAPs.

For high-T disordered case, the bond-length distribution between Cu-Cu, Cu-Au/Au-Cu, Au-Au nearest-neighbor (NN) pairs show large fluctuations. This affirms our hypothesis that, in absence of chemical SRO, the atomic-size mismatch of Cu/Au leads to local distortion that energetically stabilize the disordered phase, as also reflected in non-uniform change in bond-length in **Fig. 4a**. Introducing SRO decreased the large bond-lengths fluctuations in Cu-Au pairs. The SRO reduces the average Au bond-length by more than 1% compared to fully disordered phase. The reduced average bond lengths in SRO phase could be attributed to enhanced charge-transfer due to increased SRO.

Ye et al. [**62**] considered the atomic-size change by considering atoms are exactly at ideal lattices, and provide an empirical definition for the local lattice distortion (LLD) as, with *i*-th type of elements,

$$\text{LLD} = 100 \times \sqrt{\left[\sum_{i=1}^{n} c_i \left(1 - \frac{r_i}{\bar{r}}\right)^2\right]} \qquad \text{Eq. (5)}$$

where $\bar{r} = \sum_{i}^{n} c_i r_i$, and $c_i$ and $r_i$ are elemental composition and calculated atomic radii. Such LLD estimations in Eq. (5) are ambiguous as same atom type in different alloys can adopt different radii from changes in local chemical environment and local interactions (different electronegativity variances).

For LLD we calculate distortion from fully relaxed supercells to determine local atomic displacement:

$$\Delta u \, [\text{Å}] = \sum_{i} \sqrt{(x - x_1)^2 + (y - y_1)^2 + (z - z_1)^2} \qquad \text{Eq. (6)}$$

where (x, y, z) are relaxed positions and $(x_1, y_1, z_1)$ are ideal lattice positions [**63**]. In **Fig. 4b**, we plot the average atomic displacement ($\Delta u$) calculated using Eq. (6) for full relaxed fcc Cu$_3$Au supercell at high-T,

SRO$_1$, and SRO$_2$ cases. Here, $\Delta u$ represents the magnitude of average displacement for each atom in the supercell. Each Cu/Au atoms in disordered Cu$_3$Au have a random chemical environment that can cause lattice distortion due to difference in atomic size (1.28 and 1.44 Å) and electronegativity (1.85 and 1.92 on Allen scale).

The bond-length analysis in **Fig. 4a** shows that SRO leads to uniform Au (Cu) bond-length distribution around Cu (Au) (**Table 3**). Similar behavior was found in atomic displacement, where increasing SRO has significantly reduced $\Delta u$ variation, as shown in **Fig. 4b**. This suggests that ordering increases the chemical bonding by increasing the covalency that stabilizes the alloy as reflected through $E_f$ and change in cell volume, see **Table 2**. The bond elongation or compression based on alloying elements and their intrinsic characteristic is reflected through absolute lattice displacement. A significant decrease in the mean-squared atomic displacement in **Fig. 4b** shows that both friction stress and ability to form solid solution can be tuned using SRO, which are responsible for strength in chemically complex alloys.

Furthermore, in Fig. 5a-c we plot Au-Au, Cu-Cu, Au-Cu bond-lengths with respect to local atomic displacement ($\Delta u$) for High-T, SRO$_1$, and SRO$_2$ cases. The bond-length for High-T in **Fig. 5a** shows large spatial distribution with respect to local displacements. This arises purely from atomic-size difference of Cu and Au in absence of local chemical interaction. However, increasing SRO, i.e., lowering temperature, in **Fig. 5b, c**, shows uniform distribution of Au-Au, Au-Cu, Cu-Cu bond-lengths and displacement.

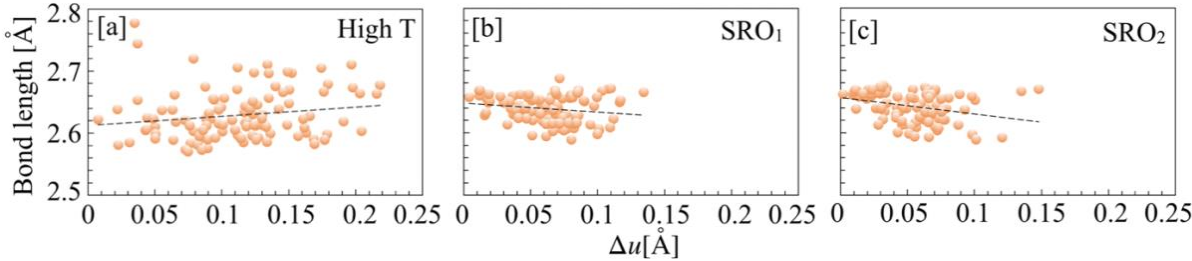

**Fig. 5**. The change in atomic bond-length and distribution for (a) High-T, (b) SRO$_1$, and (c) SRO$_2$ cases in fcc Cu$_3$Au with respect to atomic displacement ($\Delta u$).

To rationalize the properties of inorganic solids, it is important to understand the charge correlation, which could be done by studying change in valence-electron distribution. We used Mülliken and Löwdin analytical models [**46**] to assess the charge behavior in fcc Cu$_3$Au, as shown in **Fig. 6a-c**. If we carefully review the Mülliken charges for each atom in **Fig 6a**, the charge sharing between Cu-Au has significantly increased from High-T to SRO$_2$ case. Similar trends were found for Löwdin charges in **Fig. 6b**, where charge distribution has widened with increasing SRO. The increased charge sharing is commensurate with improved bonding strength with increasing SRO in **Fig. 4**. Despite some differences, we found one common thing for both Mülliken and Löwdin charge distribution that Au-gain more charge with increasing SRO, which can also be attributed both to increased SRO and higher Au electronegativity (1.92 (Au) > 1.85 (Cu) on Allen scale). Although charge spread is larger for Mülliken than Löwdin, the scatter plot in **Fig. 6c** shows similar trends, i.e., basic conclusions will not change.

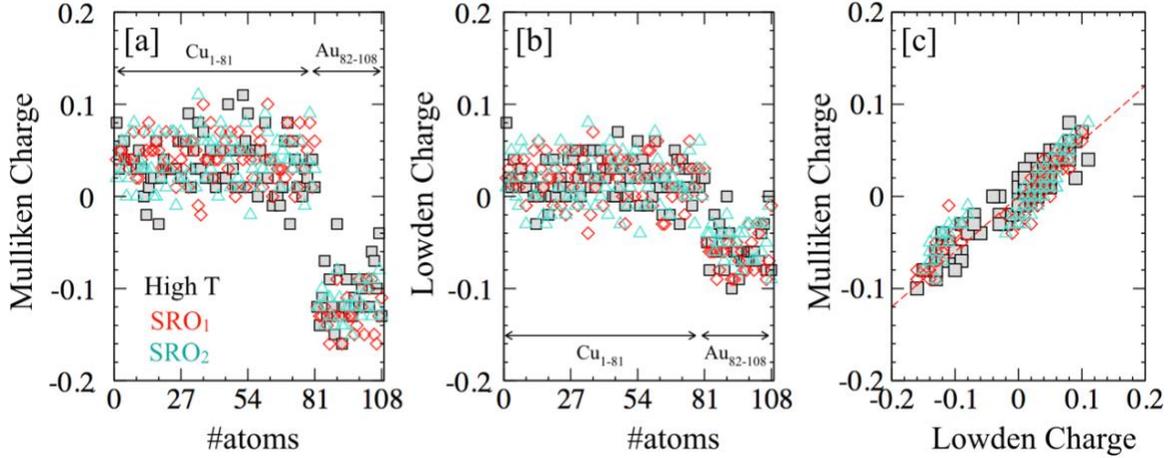

**Fig. 6**. For of fcc $Cu_3Au$, (a) Müllikan and (b) Lowden charge analysis of High-T, $SRO_1$, and $SRO_2$. (c) Comparative charge analysis of Löwden vs Müllikan.

*SRO effects on electronic structure and energy:* Here we detail the electronic structure for fcc $Cu_3Au$. The partial DOS (pDOS) for Cu and Au atoms are shown in **Fig. 7a-f** and **Fig. 8a-f**,

Comparing pDOS for Cu/Au in **Fig. 7 & 8**, the major changes in Cu-$t_{2g}$ ($d_{xy}$, $d_{xz}$, $d_{yz}$) and Au-$e_g$ ($d_{y3}$, $d_{x2-y2}$) states were observed below $E_F$ for different SRO cases, as shown in **Fig. 7b, c, e** and **Fig. 8d, f**. Clearly, the hybridization of Cu-$t_{2g}$($d_{yz}$) with Au-$s$ and Au-$t_{2g}$($d_{yz}$) have increased significantly with increase in SRO. This is also reflected in pDOS near –5 eV below $E_F$ in **Fig. 7c** and **Fig. 8a-f**. We found that Cu/Au-d pDOS at lower energies shows band-splitting and moved away from the $E_F$ with increasing SR, where the Au pDOS becomes sharper as shown for Au-*5d* in **Fig. 8**. The observed change in *d*-band smearing can be attributed to change in nearest-neighbor configurations due to SRO. Furthermore, Cu-*s* and Au-*s* states show no major changes except minor increase in pDOS at $E_F$ in **Fig. 7a** and **Fig. 8**. which correlates well with coefficient of electronic specific heat and phase stability of such alloys [**52,64**].

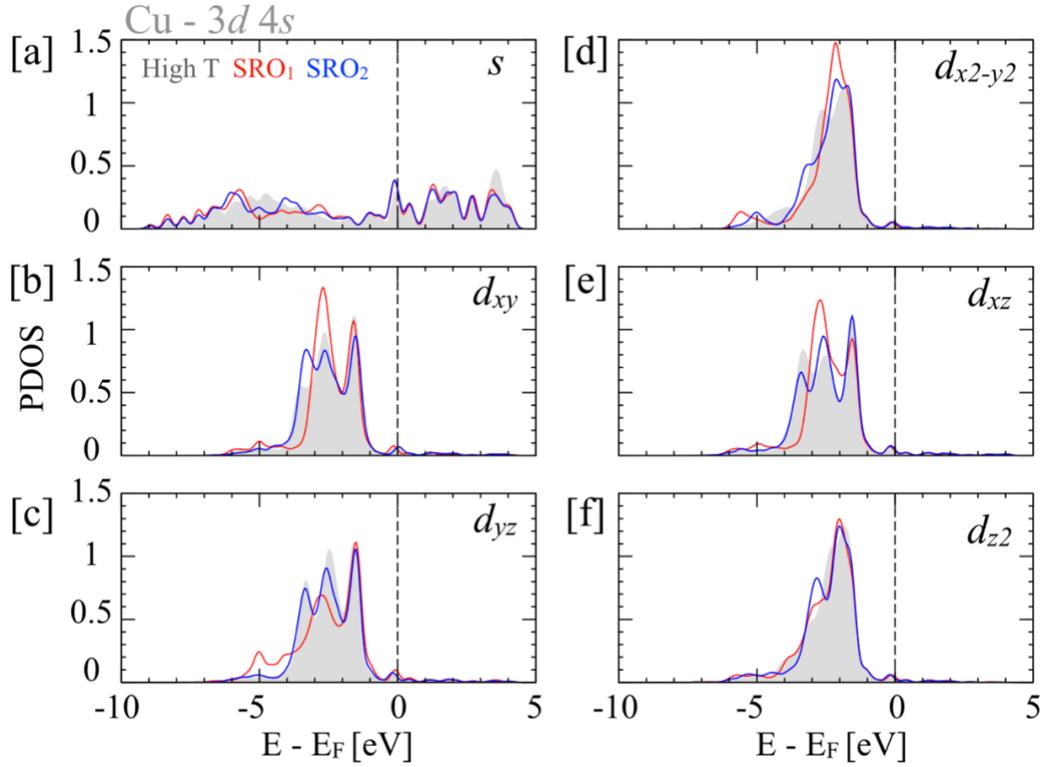

**Fig. 7**. (a-f) Cu 4*s* and 3*d* partial DOS relative to $E_F$ (vertical dashed line) for high-T (823 K), SRO$_1$ (723 K), and SRO$_2$ (678 K) of fcc Cu$_3$Au.

In disordered Cu$_3$Au, Cu/Au are mixed in 3:1 ratio, where fcc has 12 nearest neighbors (nnb), so Au and Cu will have 3 and 9 nnb, respectively. But the increased SRO modifies the configuration, so the structural and thermodynamic properties are expected to change. Moreover, in **Fig. 7** and **8**, a narrowing of Cu-*3d* bands in pDOS near –4 eV below $E_F$ is found due to increased SRO that suppresses electronic features. At high-T, disorder scattering introduces broadening due to chemical disorder, as shown by DOS (*grey*). The increased SRO sharpens the states near/at -5.25 eV but no changes were observed in states at/near -3.75 eV, which was found to improve Cu-*3d* and Au-*5d* hybridization. The major change in pDOS were seen at -4 eV below $E_F$, where Cu-*3d* states show increased hybridization with Au-*5d*. The increased Cu-Au hybridization improves the phase stability. In contrast, the relative change in electronic structure of SRO$_1$ and SRO$_2$ cases are not as significant as in high-T case. This agrees with small change in $E_f$ due to SRO, as shown in **Table 2,** where $E_f$[SRO$_1$] $-E_f$[SRO$_2$] = +0.0045 eV-per-atom. This shows that SRO can tune local electronic features and controls thermodynamic, electronic, and structural properties.

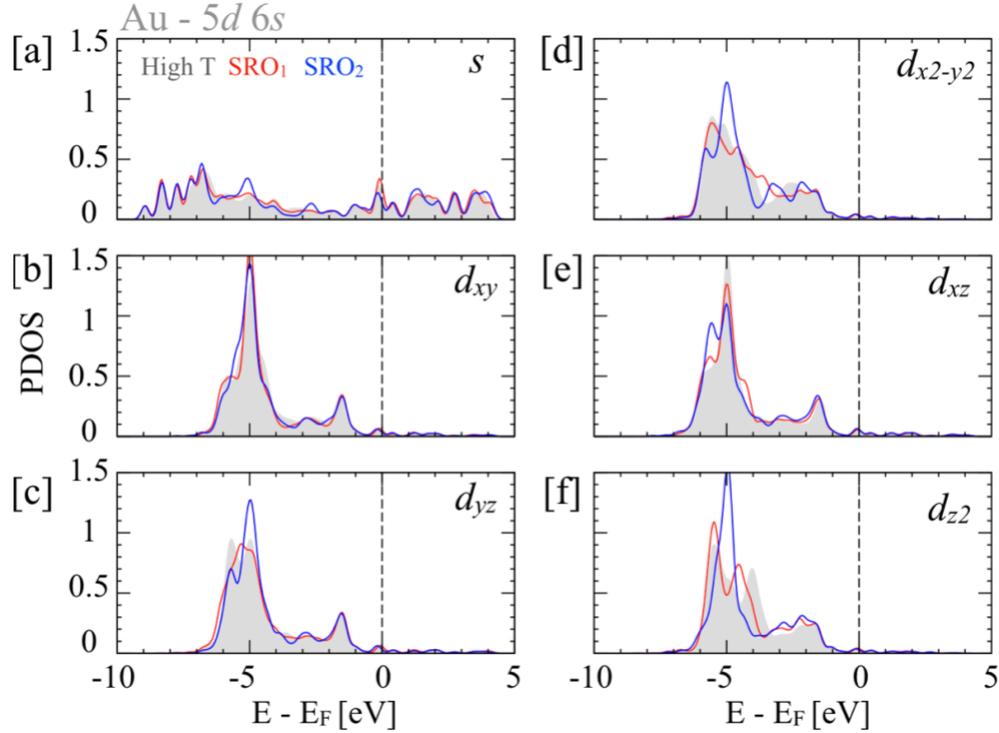

**Fig. 8**. For fcc $Cu_3Au$, (a-f) Au $6s$ and $5d$ partial DOS relative to $E_F$ (vertical dashed line) for High-T (823 K), $SRO_1$ (723 K), and $SRO_2$ (678 K).

In this section, analyzed the binding behavior of fcc $Cu_3Au$ using the crystal-orbital-Hamilton projected population. The pCOHP in **Fig. 9a,b** is indicative of bond strength in alloys. As metal-metal bonds are delocalized, we are expecting smaller charge-transfer (fewer electrons per bond) compared to covalently bonded systems (e.g., intermetallic compounds), which also depends on type of elements participating in bonding. Both Cu and Au have significant atomic size (1.28 and 1.44 Å) and electronegativity (1.85 and 1.92 on Allen scale) differences. These warrant increased charge activity with increased SRO as the same effect was observed in bonding and atomic displacement in **Fig. 5 & 6**. The charge analysis in **Fig. 6** shows that SRO significant increases the charge behavior. A positive and negative pCOHP for fcc $Cu_3Au$ in **Fig. 9a,b** indicates antibonding and bonding contributions, respectively, of local electronic states. The pCOHP shows increased number of bonding states at -5 eV. Notably anti-bonding states show significant increase near $E_F$, where the main contribution to bonding arises from deep-lying $t_{2g}$ states of Cu/Au. This behavior was also observed in NiPt, a feature responsible for the Hume-Rothery's 15% rule [**65**]. The pCOHP behavior corroborates well with increased hybridization with increasing SRO. In **Fig. 9c,d**, we show pCOBI (crystal-orbital-bond indices), which is a measure of chemical bond order. The pCOBI curve displays large increase in bonding states (negative) at higher SRO caused by an increased hybridization. This is primarily a characteristic of increased covalency.

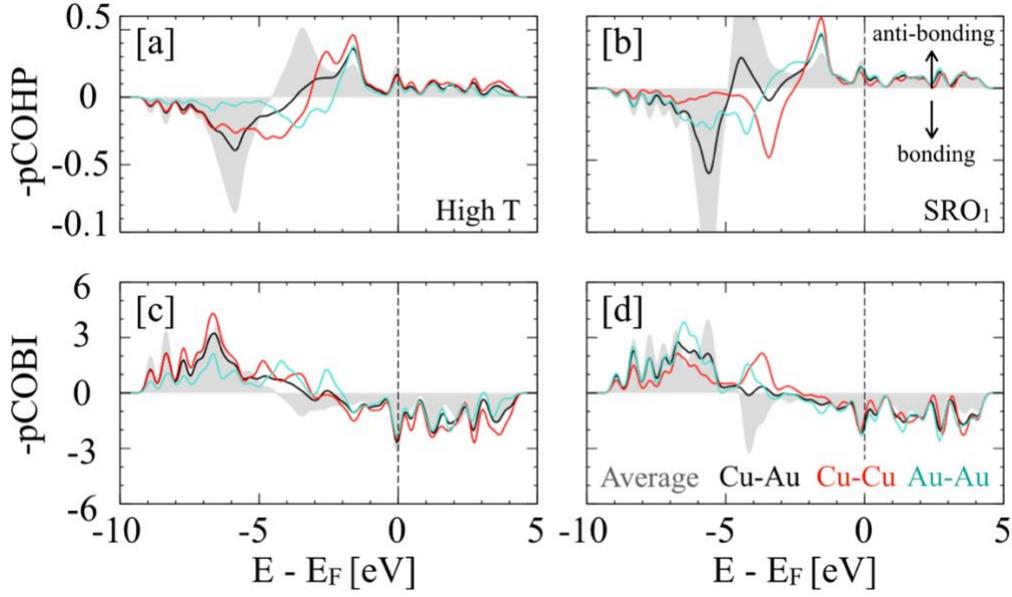

**Fig. 9.** For fcc $Cu_3Au$, (a,b) Projected crystal-orbital-Hamilton populations (pCOHP), and (c,d) projected crystal-orbital-bond indices (pCOBI) of high-T, $SRO_1$, and $SRO_2$.

*Phonon tunability:* The phonon properties of are sensitive to the force acting on atom and cell. So, we carefully relaxed each SRO-optimized SCRAPs such that net force on each atom is zero. The SRO dependent phonons were calculated at equilibrium lattice constant for fcc $Cu_3Au$ (**Fig. 10a**) in High-T (black), $SRO_1$ (red), and $SRO_2$ (blue). In **Fig. 10a**, we show an explicit effect of change in chemical correlation on shape and position of phonon modes. The phonon frequencies in **Fig. 10a** suggests that bonding behavior in disordered and SRO cases are not analogues, as is clearly visible in **Fig. 4a**.

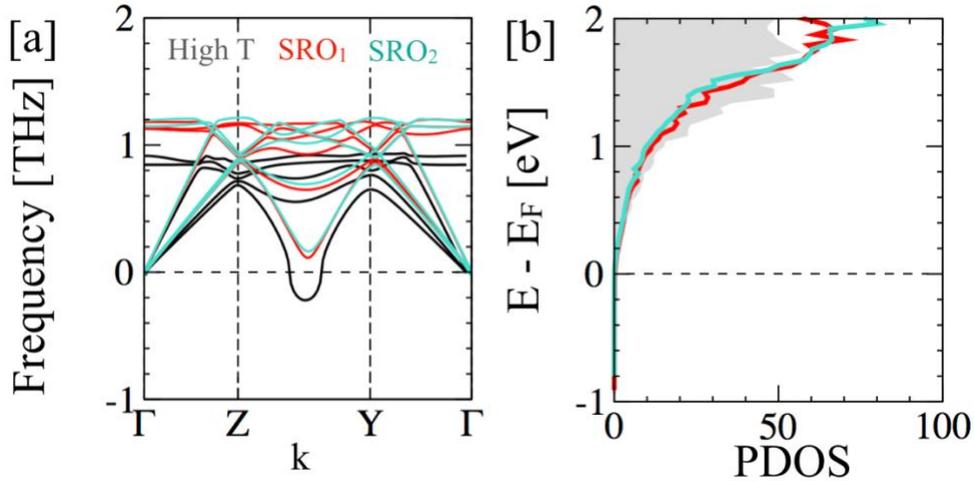

**Fig. 10**. For fcc $Cu_3Au$, the effect of increasing degree of SRO on (a) phonon dispersion (THz), and (b) phonon density of states (states/atom).

Notably, the phonon band-structure for high-T (black) case in **Fig. 10a,b** shows dynamically unstable acoustic phonon modes (imaginary frequency, here given as negative phonons) along Z-Y symmetry so does the phonon density of states. This is indicative of weak structural instability of the High-T phase of fcc Cu$_3$Au. With formation of SRO, the unstable modes are removed due to band-energy stabilization. The phonon frequency becomes more and more positive with increasing degree of SRO. The inclusion of zone boundaries shows a high density of bands in phonon band-structure coming from flat part (optical modes) of dispersion curve in **Fig. 10a**. These changes in phonon bands (or lattice vibrations) and DOS in **Fig. 10a, b** can be attributed to slow propagation of phonon wave packets and local atomic displacement ($\Delta u$) due to inclusion of SRO. We found a significant impact of SRO on $\Delta u$ (**Fig. 4b**) and bond-length stiffness (**Fig. 4a**) on phonon stability and dispersion in **Fig. 10a**. A constant shift of optical modes was also observed in the frequency range 0.75-1.0 THz to 0.8-0.9 THz 1.1-1.2 THz with increasing SRO in **Fig. 10a**, however, major changes in shape of phonon bands were seen in acoustic modes stabilized with SRO.

*SRO effect on free-energy and vibrational entropy*: In **Fig. 11**, we plot free energy and vibrational entropies for fcc Cu$_3$Au that shows noticeable change due to change in SRO. The free energy change ($\Delta F$) in SRO$_1$ and SRO$_2$ cases (**Fig. 11a**) was found to increase (i.e., more positive, see **Fig. 12**) with reduced entropic contribution ($\Delta S_{vib}$; **Fig. 11b**) relative to the High-T phase. At low T, we found a sharp increase in $\Delta S_{vib}$ that saturates nearing RT. This shows the importance of low-temperature contribution of $S_{vib}$ in phase stability. Moreover, in high-T limit, vibrational contribution saturates to a constant value between two SRO states (i.e., -1.6$k_B$ in **Fig. 11b**). This suggests that basic conclusion such phase stability and structural changes remain invariant with respect to change in $\Delta S_{vib}$. This change in phonon entropy of mixing is also reflected through small variation in phonon DOS as a function of SRO in **Fig.10b**.

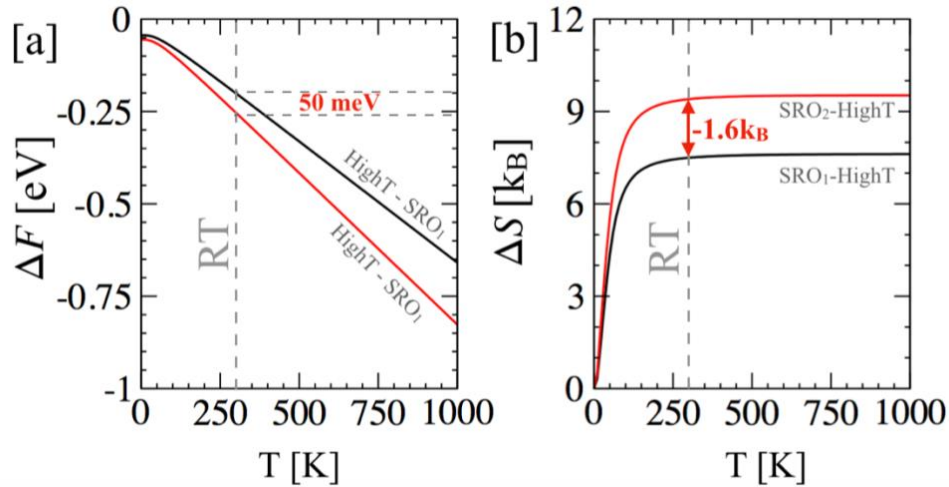

**Fig. 11**. For fcc Cu$_3$Au, the effect of SRO on (a) free-energy (eV) and (b) vibration entropy (k$_B$).

The dependence of phonon entropy of mixing on the local arrangement of atoms delivers a deeper insight into the thermodynamic stability of complex alloys. In different chemical environment, we found a large variation in free energy from 50 meV at RT to 150 meV at 1000 K due to increased degree of SRO, see **Fig. 11a**. Similarly, a maximum change of -1.6$k_B$ in $\Delta S_{vib}$ relative to disordered phase was seen in **Fig. 11b**, attributed to change in local atomic environment due to SRO. The phonon entropy provides a key detail about change in structural stability of any alloy. We show free energy (**Fig. 12a**) and vibration entropy (**Fig. 12b**) plots to provide context for **Fig. 11**. Furthermore, knowing the temperature dependence and magnitude of phonon entropy between different states of chemical complex alloys remains an issue in determining phase stability.

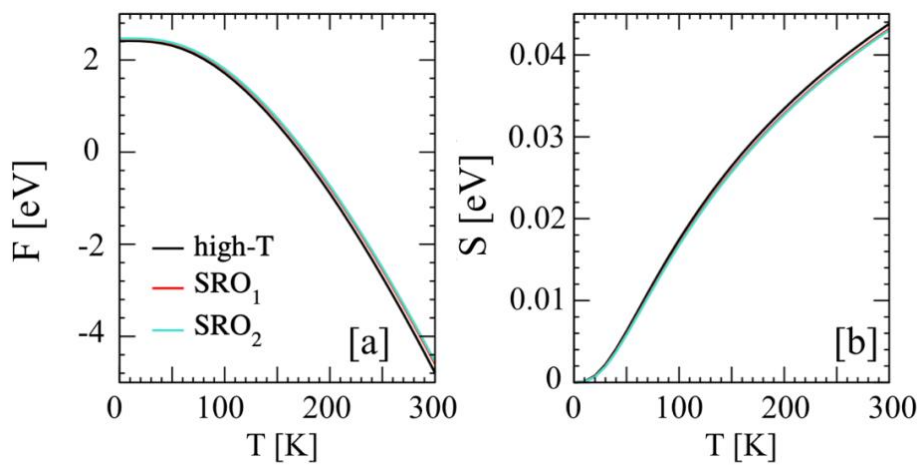

**Fig. 12**. For fcc $Cu_3Au$, (a) free-energy (kJ/mol) and (b) vibration entropy (J/K/mol) up to room temperature with increasing degree of SRO, i.e., (a) High-T, (b) $SRO_1$, and (c) $SRO_2$.

**Conclusion**

We have provided direct demonstration of the role SRO has on thermodynamic, structural, and electronic properties of archetypal fcc $Cu_3Au$. We critically analyzed the DFT-derived thermodynamic properties and short-range order, and its connection to the underlying electronic origins. The thermodynamics linear-response theory was used for quantitative analysis of SRO and estimate of the low-temperature ordering modes (e.g., $L1_2$) in fcc $Cu_3Au$, The DFT-SRO method accurately predicted the order-disorder temperature and Warren-Cowley SRO parameters for $Cu_3Au$, in agreement with experiments. We show that degree of SRO has strong effect on phase stability (formation energy at 0 K) and structural stability (phonons) in fcc $Cu_3Au$. The improved stability was also reflected in enhanced Cu-Au bonding strength and hybridization. To provide a clear bonding picture, a detailed bonding analysis was done using projected crystal-orbital-Hamilton populations method and show a direct correlation between degree of SRO and bonding strength. The effect of SRO was also seen on overall electronic structure, atomic charges, covalency and vibrational entropy of fcc $Cu_3Au$. The significant increase in bonding strength due to increased SRO helped remove the structural instability. Our results are examples of how SRO can be used as a control parameter to tune the material properties. Importantly, the methods described here are applicable to any arbitrary solid-solution alloy, as found in our recent works, including multi-principal-element alloys with and without vacancies.

**Acknowledgements**

Will Morris is grateful for the research opportunity at Ames National Laboratory supported by the U.S. Department of Energy (DOE), Office of Science, Science Undergraduate Laboratory Internships (SULI) program. Theory advances for modeling short-range order in alloys at Ames National Laboratory was supported by U.S. DOE, Office of Science, Basic Energy Sciences, Materials Science and Engineering Department. Ames National Laboratory is operated by Iowa State University for the U.S. DOE under Contract No. DE-AC02-07CH11358.

**CRediT roles**

**W.M.**: Data curation; Formal analysis; Funding acquisition; Investigation; Visualization; Roles/Writing - original draft. **D.D.J.**: Methodology; Resources; Formal Analysis; Funding acquisition; Writing - review &

editing. **P.S.**: Conceptualization; Data curation; Formal analysis; Funding acquisition; Investigation; Methodology; Resources; Supervision; Visualization; Writing - original draft; Writing - review & editing.